\documentclass[]{jdsart}
\usepackage{setspace}
 
\volume{0}
\issue{0}
\pubyear{2025}
\articletype{research-article}
\doi{0000}

\usepackage{siunitx} 
\graphicspath{{.}{./images}}

\usepackage{xcolor}

\usepackage{comment}
\usepackage{booktabs, textgreek}
\usepackage{subcaption}
\usepackage{enumitem}
\usepackage{tikz} 
\usepackage{hyperref}   
\usepackage{url}        
\usetikzlibrary{positioning} 

\begin{document}



\begin{frontmatter}
  
\title{Principles for Open Data Curation: A Case Study with the New
York City 311 Service Request Data}
\runtitle{Principles for Open Data Curation}

\author[1]{\inits{D.}\fnms{David}~\snm{Tussey}}
\author[2]{\inits{J.}\fnms{Jun}~\snm{Yan}}
\address[1]{\institution{NYC DoITT}, \cny{USA}}
\address[2]{Department of Statistics,
  \institution{University of Connecticut}, \cny{USA}}

\hyphenpenalty=950

\begin{abstract}
In the early 21st century, the open data movement began to transform 
societies and governments by promoting transparency,
innovation, and public engagement. The City of New York (NYC) has been at
the forefront of this movement since the enactment of the Open 
Data Law in 2012, creating the NYC Open Data
portal. The portal currently hosts 2,700 datasets,
serving as a crucial resource for research across various domains, 
including health, urban development, and transportation. However, the
effective use of open data relies heavily on data quality and
usability, challenges that remain insufficiently addressed in the
literature. This paper examines these challenges via a
case study of the NYC 311 Service Request dataset,  identifying key
issues in data validity, consistency, and curation efficiency. We
propose a set of data curation principles, tailored for
government-released open data, to address these challenges.
Our findings highlight the importance of harmonized field definitions,
streamlined storage, and automated quality checks, offering practical
guidelines for improving the reliability and utility of open datasets.
\end{abstract}


\end{frontmatter}

\section{Introduction} 
\label{sec:intro}

In the early 21st century, the open data movement began 
to take shape, driven by the fundamental belief that 
freely accessible data can transform both societies and 
governments. This movement champions the principles
of transparency, innovation, and public engagement. 
A landmark in this journey was the launch of the United States'
Data.gov portal in 2009 \citep{dataGov}, a pioneering
platform in making government data widely accessible. Shortly afterwards,
the European Union followed suit, unveiling its Open Data Portal 
\citep{dataEU} in 2012, further cementing the movement's 
global reach. Furthermore, the World Bank's Open
Data initiative, initiated in 2010, stands out as a comprehensive
repository for global development data, available at
World Bank Open Data \citep{dataWorldBank}. 
These initiatives represent significant strides in democratizing data, 
removing barriers that once kept valuable information 
regarding government performance in silos. Their collective impact 
extends beyond mere data sharing to fostering a culture of openness 
that benefits individuals, communities, governments, and economies worldwide 
\citep{barns2016mine, wang2016adoption}.
Despite these global efforts, the effective use of open data hinges on
addressing critical challenges in data curation, such as ensuring
quality, consistency, and usability.

The City of New York (NYC) has emerged as a leader in the open data movement,
marked by the enactment of the Open Data Law in 2012
\citep{zuiderwijk2014open}. This landmark legislation led to the
creation of the NYC Open Data Portal \citep{dataNYC}, which 
today hosts an impressive array of 2,700 datasets
across 80 different city agencies. This wealth of data serves as a
powerful tool for researchers and policymakers, significantly
enhancing local government transparency.
Popular datasets include information on
restaurant health inspection violations, car crashes, high school and
college enrollment statistics, jail inmate charges, and the location
of City-wide free Internet access points. These datasets have been
applied in civil life in various ways, such as mapping car crashes
involving pedestrians and visualizing high school and college
enrollment trends. Furthermore, they have enabled significant research
across multiple domains, including health \citep{cantor2018facets, 
shankar2021data}, urban development \citep{neves2020impacts}, and
transportation \citep{gerte2019understanding}, aiding in the
understanding and addressing of complex urban challenges.
For example, NYC’s 311 data has been used to optimize resource
allocation in urban planning and improve emergency response times,
showcasing its practical relevance in real-time decision-making.

However, the open data presents substantial curation challenges.
Data curation, the process of organizing, maintaining, and ensuring
the quality of datasets, plays a crucial role in maximizing the
utility of open data. Proper curation ensures that datasets remain
consistent, accurate, and useful for diverse applications. For example, 
well-curated data is essential for machine learning systems, which 
require high-quality data to produce reliable insights 
\citep{polyzotis2019data, jain2020overview}. A critical component of 
data curation is data cleaning, which involves identifying and rectifying 
inconsistencies, errors, and inaccuracies in datasets. While open data 
initiatives have made vast amounts of data available, ensuring the 
reliability and utility of such data hinges on rigorous data cleaning processes. 
Efficient usage of software tools can significantly streamline this 
aspect of curation \citep[e.g.,][]{cody2017cody, van2018statistical}. Poor 
curation may result in issues such as missing data, formatting errors, 
or inconsistent values, leading to biased or inaccurate outcomes 
\citep{geiger2020garbage}. This is particularly critical in domains 
where machine learning is applied to sensitive tasks, such as public 
health or policy \citep{rahm2000data}.

Research into data curation has explored these challenges in-depth. 
Among the earliest discussions, \citet{witt2009constructing} focused 
on developing data curation profiles tailored to specific contexts, 
setting a precedent for targeted data management strategies. 
Addressing broader challenges in data sharing and management, 
\citet{borgman2012conundrum} highlighted the complexities of 
research data distribution, emphasizing the need for robust 
strategies. This is complemented by \citet{hart2016ten}, who outlined 
essential principles for effective data management, particularly 
emphasizing meticulous curation practices. The utility of curated 
open data is vividly illustrated in public health and global challenges, 
where \citet{cantor2018facets} demonstrated the utility of curated 
data in evaluating community health determinants, and 
\citet{shankar2021data} observed its critical role during the 
COVID-19 pandemic in managing collective responses.
Despite these advances, there remains a notable gap in providing
actionable frameworks for managing large-scale municipal datasets.

By addressing this gap, this study transitions from theory to
practical insights, aiming to establish principles that enhance both
the reliability and utility of open government data.
The contributions of this paper are twofold. First, we delve into
the specifics of data curation challenges using the NYC 311 Service
Request (SR) Data as a case study. This dataset serves as a prime 
example for examining key issues in data curation, including data 
validity, consistency, and curation efficiency. We illustrate these 
points with live examples drawn from our 
processing of the 311 SR data. Secondly, building upon insights 
gained from this case study, we propose a set of data curation 
principles tailored for government-released open data. These 
principles are designed to address the unique challenges 
and requirements observed in the curation of such datasets.

The paper is organized as follows. Section~\ref{sec:data} provides 
an overview of the history of the NYC 311 SR system and presents 
a summary of SR counts over a 2-year period from 2022 to 2023. 
Section~\ref{sec:issues} delves into specific data cleansing 
challenges affecting data quality and curation efficiency, 
including structural problems, adherence to the data dictionary, 
and the presence of missing, blank, or N/A entries. This section
 also investigates data field compliance with reference or acceptable values, 
highlights logical inconsistencies, and examines some concerning patterns 
in the data. We explore the balance between precision and accuracy 
and identify duplicate or redundant fields, along with observations 
on the Data Dictionary \citep{datadictionaryNYC}.
Section~\ref{sec:recommendations} offers 
practical recommendations for mitigating or resolving these issues, 
while Section~\ref{sec:discussion} encapsulates key insights and 
discusses the broader implications of our findings.

\section{NYC 311 Service Request (SR) Data} 
\label{sec:data}

The NYC 311 service, a critical component of New York City's public
engagement and service response framework, serves as a centralized hub
for non-emergency inquiries and requests. Introduced in 2003, the NYC
311 system was designed to streamline the city's response to
non-emergency issues, ranging from noise complaints to street
maintenance requests. Initially a phone-based call center, the system
evolved into a comprehensive data management platform handling
millions of requests annually. Key milestones since its launch in 2003
include the addition of online and mobile channels in 2009, a
record monthly high of 348,463 SRs in 
August 2020 (associated with the the COVID-19 
pandemic), the 2021 expansion to include the 
Metropolitan Transit Authority (MTA) NYC subway 
system (the biggest expansion in 311 history), 
and a yearly record of 3.23 million SRs in 2023.

Today, the NYC 311 data system manages over 3 million service
requests (SRs) per year. This data is publicly accessible 
through the NYC Open Data Portal, which provides 
tools for querying, grouping, aggregating,
visualizing, and exporting results. The NYC Office of Technology and Innovation (OTI) 
provides technical support for the 311 application, 
the open data infrastructure, mobile \& web applications, 
and the public-facing data portal. Input data for the 311 system 
is sourced from 16 different NYC Agencies. Some of these Agencies
utilize the core 311 software, while others transfer data to the
311 Open Data Warehouse from their in-house systems.

The impact of NYC 311 data extends beyond operational efficiency; it
has become instrumental in shaping City governance and community
engagement. Open data not only ensures governmental transparency
but also empowers civic developers, the general public, and
policy-makers \citep{minkoff2016nyc, o2017uncharted,
  kontokosta2021bias}.
The data has been pivotal in providing advice on shelters
during emergencies, handling inquiries during the COVID-19 pandemic,
enforcing standards between landlords and tenants, reallocating taxi
routes based on analyses by the Taxi and Limousine Commission (TLC),
and improving responsiveness across City Agencies.
It also supports a wide range of studies in resource allocation and
emergency response strategies \citep{zha2014profiling, raj2021swift},
social equity in service provision \citep{white2018promises,
  kontokosta2021bias}, and urban challenges such as noise pollution
\citep{dove2022sounds} and street flooding
\citep{agonafir2022understanding}.

Our investigation uses a 311 SR dataset covering  2-years (2022-2023), 
and was collected on 15 September 2024. It is to conduct 
analysis in areas such as data consistency, data validity, 
redundant fields, etc.. Instructions on downloading this file are 
available in the supplemental material. Characteristics of this
two-year 311 SR dataset are as follows.

\begin{itemize}[left=1.5em]
\item The dataset is 3.7 GB in size and contains 
  over 6.4 million rows. It exports as a CSV file.

\item There are 41 columns (fields) of data for each row, with each
  row representing a single SR. To facilitate 
  processing, column names were made lower case and 
  spaces were replaced with underscores, e.g. \texttt{complaint\_type}. 
  This naming convention is used throughout this paper. 
	
\item Each SR has four date fields: 
	\begin{itemize}	
		\item{\texttt{created\_date}}
		\item{\texttt{closed\_date}}
		\item{\texttt{resolution\_action\_updated\_date}}
		\item{\texttt{due\_date}}
	\end{itemize}

\item There are two borough fields, \texttt{borough} \& \texttt{park\_borough}, 
		which appear to be duplicates.
  
\item There are seven street fields; two pair of which appear to be duplicates:
	\begin{itemize}
		\item{\texttt{incident\_address}}
		\item{\texttt{street\_name}}
		\item{\texttt{cross\_street\_1}}
		\item{\texttt{cross\_street\_2}}
		\item{\texttt{intersection\_street\_1}}
		\item{\texttt{intersection\_street\_2}}
		\item{\texttt{landmark}}
	\end{itemize}
          
\item In addtion to \texttt{incident\_address}, each SR has eight
	additional geographic fields:
	\begin{itemize}
		\item{\texttt{latitude} \& \texttt{longitude}}
		\item{\texttt{location}}
		\item{\texttt{street\_name}}
		\item{\texttt{landmark}}
		\item{\texttt{block}}
		\item{\texttt{x\_coordinate\_state\_plane} \& \texttt{y\_coordinate\_state\_plane}}.
 	\end{itemize}
	
\item A free-form text field, \texttt{resolution\_description}, which 
  supports 930 characters including commas and 
  special characters; often problematic for automated processing.
\end{itemize}

During the 10-year period from 2014 to 2023,  NYC's 311 system
experienced a 50\% increase in the number of SRs, 
reflecting a growing public reliance on and usage of 
the 311 system. (During this 10-year time frame, the 
population of NYC grew just 0.6\%.). The most significant increase
in activity occurred in 2020, when the volume of requests 
surged to 3.23 million during the COVID-19 pandemic, 
illustrating the essential role of the 311 system in supporting 
NYC residents during times of crisis. This steady 
rise in SRs can be at least partially attributed to the increased 
accessibility of the 311 system via online and mobile 
platforms, as well as heightened public awareness of the service. 
While the spike in 2020 was exceptional, the broader trend 
indicates sustained growth in SR volume in future years,  
highlighting the system's expanding role in managing 
both routine City operations and extraordinary events.

\begin{figure}[tbp]
	\centering
	\includegraphics[width = \textwidth]{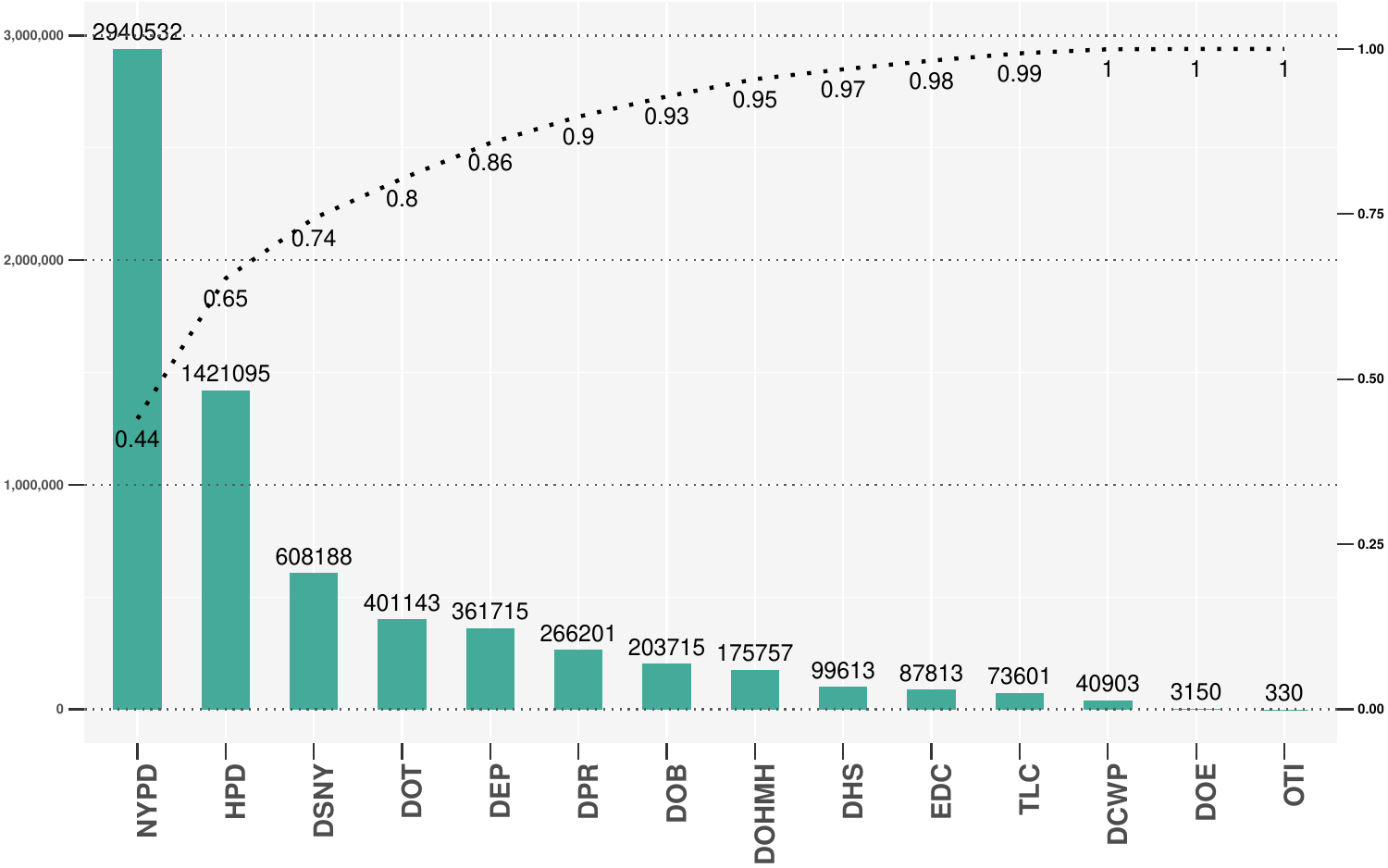}
  	\caption{311 SR counts by agency for the time period 2022--2023.}
	\label{fig:SRcountbyAgency}
\end{figure}

Figure~\ref{fig:SRcountbyAgency} provides a breakdown of SRs by
the responsible Agency, showing the cumulative 
percentage of SRs handled by each
agency over the 2-year time frame. Note that the
distribution of SRs is heavily concentrated among a few key 
agencies. The six largest agencies are:

\begin{enumerate}[left=1.5em]
    \item New York Police Department (NYPD): 43\% of all SRs
    \item Housing Preservation and Development (HPD): 21\%
    \item New York City Department of Sanitation (DSNY): 10\%
    \item Department of Transportation (DOT): 7\%
    \item Department of Environmental Protection (DEP): 5\%
    \item Department of Parks and Recreation (DPR): 4\%
\end{enumerate}

Collectively these six City Agencies handle over 90\% of the total SR 
volume, indicative of the critical role they play in managing
public concerns; ranging from noise complaints and housing issues to
sanitation and transportation issues. The remaining
10\% of SRs are distributed across 10 additional Agencies. 
This concentration of SRs to these ``big six'' Agencies 
underscores the necessity for optimal data curation processes 
within these  high volume Agencies.

\begin{figure}[tbp]
 \centering
  \includegraphics[width = \textwidth]{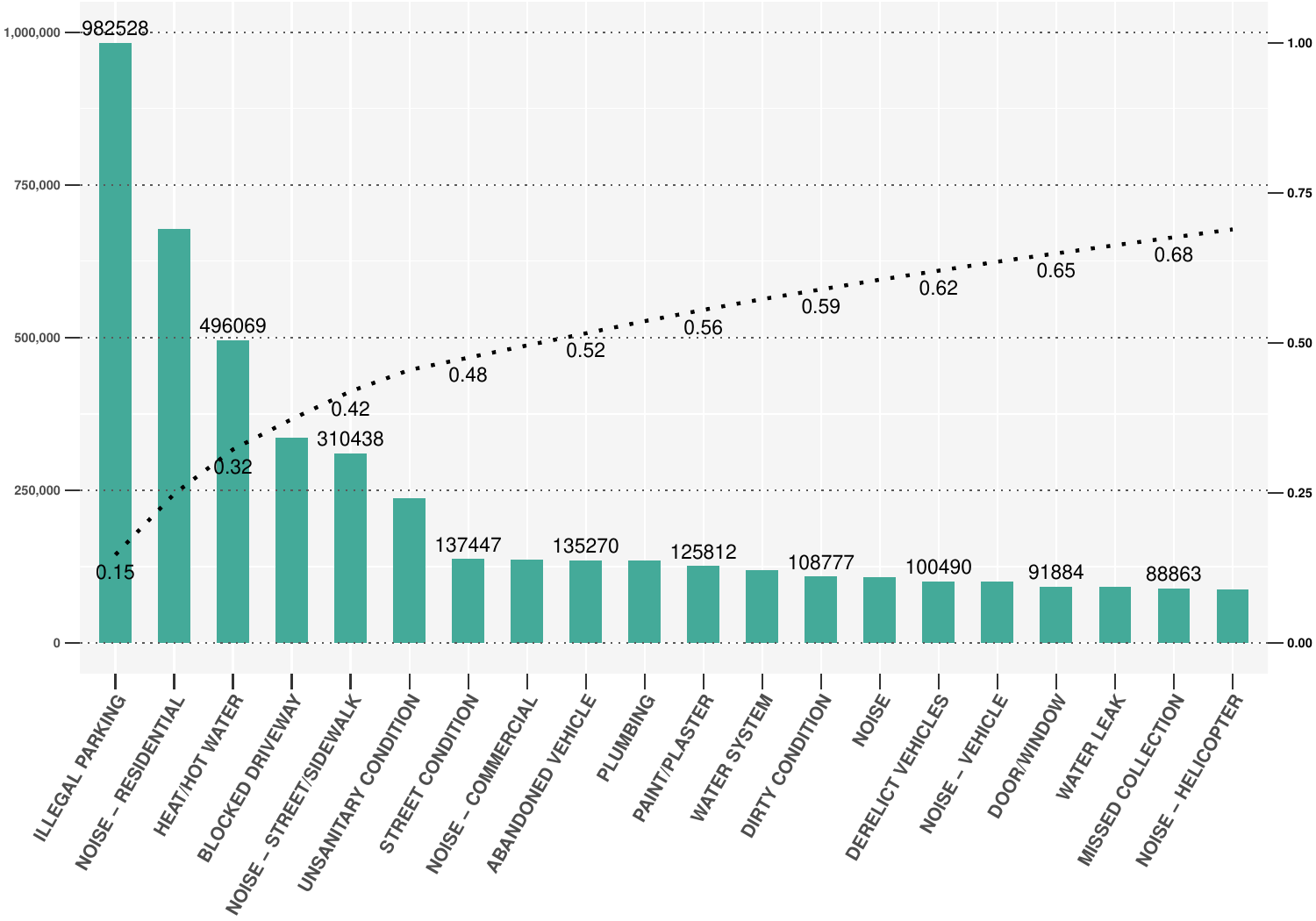} 
  \caption{Top 20 complaint types for the 311 SR during 2022--2023.} 
  \label{fig:SR_complaints}
\end{figure}

The dataset contains 210 different \texttt{complaint\_types}. 
Figure~\ref{fig:SR_complaints} illustrates how the Top 20 
\texttt{complaint\_type(s)} account for 70\% 
of all SRs. The distribution of complaints is skewed, 
with a small number of issues dominating total SR volume. 
Of special interest are noise-related complaints, of 
which there are eight different types including vehicle, 
residential, commercial, and street noise. When 
combined, noise-related SRs make up 22\% of all 
complaints; the most frequent issue in the NYC 311 system. 
Other prominent categories include illegal parking, heat/hot water 
complaints, and blocked driveways. The cumulative percentage curve 
reveals that the Top 20 \texttt{complaint\_type(s)} comprise 70\% of SRs,
while the remaining 190 other \texttt{complaint\_type(s)}  are spread thinly 
across the remaining 30\%. This suggests that improving 
responses to the most frequent complaints could have an 
outsized impact on overall service efficiency and resident 
satisfaction. This chart also provides insights into the 
operational pressures faced by those City Agencies 
responsible for handling these high-volume complaints.

\section{Data Cleansing Issues} 
\label{sec:issues}
Data cleansing refers to the process of identifying and rectifying
errors, inconsistencies, and inaccuracies within datasets to ensure
they are of high quality and reliable for analysis
\citep{maletic2005data, hosseinzadeh2023data}. The process
typically involves removing duplicate records, handling missing or
incomplete data, correcting mislabeled or inaccurate entries, and
standardizing data formats \citep[e.g.,][]{cody2017cody,
  van2018statistical}. In the context of open data, cleansing is
especially important as open datasets often come from diverse,
often uncoordinated sources, leading to variations in data quality,
completeness, and consistency. Without thorough data cleansing, 
the utility of open data can be limited and perhaps 
untrustworthy, affecting its reliability for research, 
policy making, and innovation. The main purpose of cleansing 
open data is to ensure that it is accurate, consistent, and usable a
cross multiple analytical platforms and by various stakeholders with a 
myriad of purposes. Data cleansing improves the trustworthiness 
of the data and enables more accurate analysis, 
better decision making, and the improved integration of 
data into machine learning models or other systems.

Many quality criteria are employed to ensure high-quality data 
processing. One of the primary efforts is data validation, which includes 
several critical checks. For instance, mandatory fields must not be 
left empty, ensuring that key information is always captured. 
Additionally, certain fields must conform to specific data types, 
such as numeric, character, or date formats; typically 
outlined in a Data Dictionary. An essential aspect of 
validation is domain compliance, where 
fields must adhere to a predefined set of 
values such as statuses, state names, or zip codes. Structural 
errors can also play a significant role in data quality, particularly when 
naming conventions or data entries are inconsistent. Common issues 
include fields that do not appear in the Data Dictionary and  
inconsistent usage of data fields. Furthermore, redundant 
(or irrelevant) fields can clutter datasets, reducing efficiency, 
creating confusion, and introducing errors. Logical inconsistencies 
are another consideration, such as related fields that violate expected 
relationships, e.g. a ``due date'' that precedes the ``created date.'' 
Lastly, the balance between accuracy and precision is crucial, as both 
must be managed to ensure meaningful data.

We identify the presence of issues in the NYC 311 SR open dataset;
not attempting to solve them, but rather to highlight their scope, 
magnitude, and potential impact. We would recommend 
that a corrective solution should be undertaken only 
after further investigation as to the why and how such 
issues might come about, including  detailed discussions 
with the originating agency as to whether or not the issue is an 
actual error, correct,  or holds some other status.

\subsection{Structural Issues}
\label{sec:structural}
Structural issues refers to how data is organized, formatted, 
or presented within a dataset. Structural issues can cause difficulty in 
analyzing the data effectively, often necessitating extensive 
data cleaning. We identified three primary structural issues in the
NYC 311 dataset: undocumented fields, inconsistent representations of
missing data, and daylight saving time anomalies.
 
\paragraph{Undocumented Fields:} The 311 SR 
The 311 SR Data Dictionary identifies 
41 data columns (fields) and provides related information 
for each. However, there are six additional fields that can be
optionally downloaded if manually selected. These fields are 
visible in the portal's ``Column manager'' widget, labeled with the prefix 
``@computed\_region.'' and include:

\begin{itemize}[left=1.5em]
    \item \texttt{zip\_codes}
    \item \texttt{community\_districts}
    \item \texttt{borough\_boundaries}
    \item \texttt{city\_council\_districts}
    \item \texttt{police\_precincts}
    \item \texttt{police\_precinct}
\end{itemize}

While one can likely infer the meaning of these \emph{computed} 
fields, uncertainty remains as to their derivation and 
usage. These six fields are not included in
the 311 SR Data Dictionary. Additionally, our analysis 
revealed that these six ``computed'' fields have 
significant data validity issues, leading us to categorize these six fields 
as experimental and \emph{not for public use}. An update 
to the Data Dictionary would seem appropriate. Accordingly, 
we omit these fields in our analysis. 

\paragraph{Representation of Missing data:} We found that the 
311 SR dataset exhibited inconsistent approaches when 
presenting missing data. We discovered different 
representations of missing data: nulls, spaces, ``\texttt{NA}'', ``\texttt{N/A}'', 
and ``\texttt{<NA>}''. This variety of representations complicates programming
efforts and increases the likelihood of introducing errors 
in data analysis (typically when selecting data for blank, NA, or missing fields). A more consistent 
approach to treating missing data should be employed to address this issue.

\paragraph{Challenges with Daylight Saving Time (DST):} The 311 SR 
dataset captures local NYC times. Accordingly, twice each 
year there are changes to the local time as a result of observing 
DST. This can introduce some unexpected results. For example, 
on both March 13, 2022 and March 12, 2023 there are no times 
between 1:59AM and 2:59AM. The 2:00AM hour on those dates 
is simply missing owing to the clocks moving forward; at 2:00AM 
it instantly becomes 3:00AM. This creates a gap in the hourly 
distribution of SRs on those days, and exaggerates the 
lifespan of SRs, albeit by a single hour. Similarly on both
November 6, 2022 and November 5, 2023, the 
hour of 1:00AM - 2:00AM is repeated twice. This can result 
in a bizarre scenario where, for example, an SR can 
be \emph{created} at 1:59AM and then be \emph{closed} 30 minutes later 
at 1:29AM; seemingly nonsensical. Utilizing UTC time 
for date-time stamps would help address this issue.

\subsection{Missing Data by Field}
\label{sec:blanks}
Understanding the absence of data by field greatly aids analysis. 
For example, determining if SRs were closed 
before their \texttt{due\_date}, would be challenging as 99.6\% of the
\texttt{due\_date} entries are blank as only DSNY appears to use this field.
Blank (or N/A values) divide into three groups:
\begin{itemize}[left=1.5em]
    \item Mostly Empty: 93-99.9\% blank;
    \item Partially Empty: 40-4\% blank;
    \item Few/None Empty: 2-0\% blank.
\end{itemize}

The Mostly Empty category includes:
\texttt{taxi\_company\_borough}, 
\texttt{due\_date}, 
\texttt{road\_ramp},
and \texttt{bridge\_highway\_name}. 
The Partially Empty category includes:
\texttt{location\_type}, 
\texttt{landmark} and \texttt{cross\_street\_1 \& \_2}.
The Few/None Empty category includes \texttt{created\_date}, 
\texttt{complaint\_type},
\texttt{agency}, 
and \texttt{status}. 
Note that several of the Mostly Empty fields are used by only one or 
two Agencies (See: \autoref{tab:field-usage-summary}).

\begin{table}[tbp]
    \centering
    \caption{Fields used by only one or two Agencies with limited non-blank entires}
    \label{tab:field-usage-summary}
    \small
    \begin{tabular}{l r r p{4cm}}
        \toprule
        \textbf{Field} & \textbf{\# of entries} & \textbf{\% non-blank} & \textbf{Agencies using} \\
        \midrule
        \texttt{taxi\_company\_borough} & 3,567  & 0.06\% & TLC \\
        \texttt{vehicle\_type}           & 18,642 & 0.29\% & TLC, NYPD \\
        \texttt{due\_date}               & 24,414 & 0.38\% & DSNY \\
        \texttt{taxi\_pick\_up\_location} & 64,954 & 1.02\% & TLC \\
        \bottomrule
    \end{tabular}
\end{table}

\begin{figure}[tbp]
	\centering
  	\includegraphics[width=\textwidth]{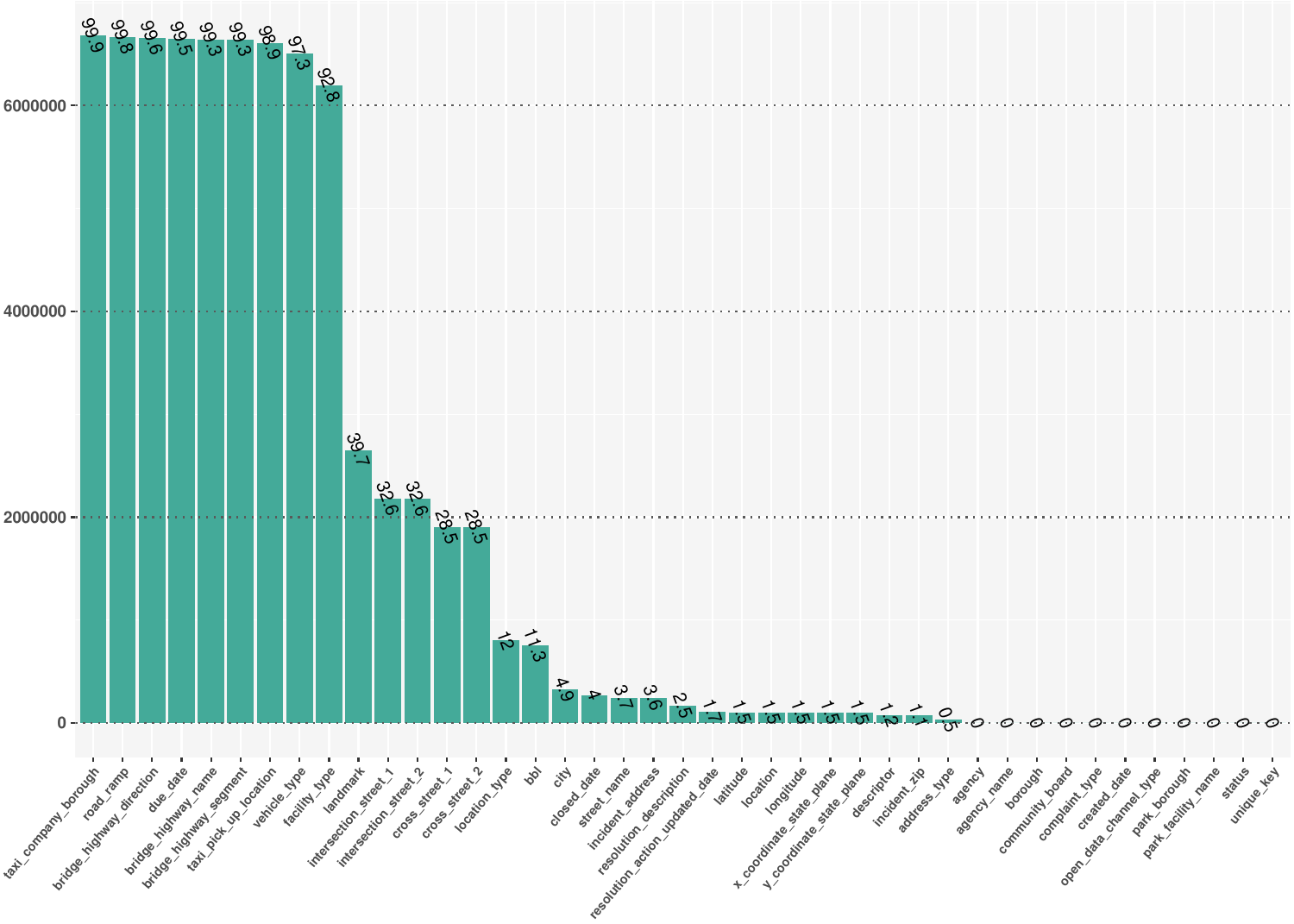}
	\caption{Number and percentage of empty/blank entries for the
          311 SR data during 2022--2023.}
	\label{fig:blank_fields}
\end{figure}

For some analysis efforts, it may be prudent to inquire 
as to why some fields are mostly/almost always blank. 
Figure~\ref{fig:blank_fields} is a graphic depiction of the percentage of
empty (blank \& N/As) for each field. It illustrates the grouping into 
Mostly Empty, Partially Empty, and Few/None Empty segments.

Table~\ref{tab:field-usage-by-agency} highlights the varied usage of
certain fields across NYC agencies, revealing both specialization and
redundancy. For instance, \texttt{landmark} and
\texttt{facility\_type} are widely used, indicating their importance
in classifying SRs. In contrast, fields like
\texttt{taxi\_company\_borough} and \texttt{vehicle\_type} are used
exclusively by TLC, suggesting that such fields, which serve a narrow
purpose, could be better stored in separate, agency-specific tables.
This approach would reduce the size and complexity of the primary
dataset, making it easier to maintain and analyze while preserving
the ability to perform specialized analyses when necessary.
Streamlining through harmonization and validation would further
enhance consistency and optimize cross-agency analyses.

\begin{table}[tbp]
\centering
\caption{Sample field usage by select agencies among the 6.4 million
  SRs in 2022--2023. }
\label{tab:field-usage-by-agency}
\small
\begin{tabular}{lrrrrrrr}
\toprule
\textbf{\textbf{Field}} & \textbf{DEP} & \textbf{DHS} & \textbf{DOB} & \textbf{DOT} & \textbf{DPR} & \textbf{DSNY} & \textbf{TLC} \\
\midrule
\texttt{landmark}                  & 2935     & 70185    & 282      & 135150   & 254218   & 424284   & 63133  \\
\texttt{facility\_type}            & 175549   & 0        & 51575    & 159667   & 0        & 66012    & 0      \\
\texttt{due\_date}                 & 0        & 0        & 0        & 0        & 0        & 32097    & 0      \\
\texttt{vehicle\_type}             & 0        & 0        & 0        & 0        & 0        & 0        & 1696   \\
\texttt{taxi\_company\_borough}    & 0        & 0        & 0        & 0        & 0        & 0        & 4025   \\
\texttt{taxi\_pick\_up\_location}  & 0        & 0        & 0        & 0        & 0        & 0        & 72558  \\
\texttt{bridge\_highway\_name}     & 15       & 21012    & 0        & 7451     & 0        & 613      & 977    \\
\texttt{road\_ramp}                & 15       & 2588     & 0        & 7450     & 0        & 613      & 999    \\
\bottomrule
\end{tabular}
\end{table}

\subsection{Validating Data for Acceptable Values}
\label{sec:domain}
Any analytic effort must ensure that fields containing invalid values 
are identified and isolated from the analysis. For example, the \texttt{latitude} 
and \texttt{longitude} fields were found to all fall within the 
geographic boundaries of New York City. Likewise, 
the \texttt{unique\_key} field was indeed unique, as required. 
Unfortunately, the 311 SR Data Dictionary specifies very few domains
of acceptable values for other fields. Nonetheless, the following
fields were tested and found to comply with their expected
domains, as determined by public
usage and as evidenced in other historical datasets:
\begin{itemize}[left=1.5em]
    \item \texttt{address\_type}
    \item \texttt{status}
    \item \texttt{borough}
    \item \texttt{borough\_boundaries}
    \item \texttt{park\_borough}
    \item \texttt{data\_channel}
    \item \texttt{vehicle\_type}
    \item \texttt{city\_council\_district}
\end{itemize}
However, some fields lacked compliance
to a domain of allowable values, as outlined next.

\paragraph{Zip Codes:}
\label{sec:zipcodesissues}
All \texttt{incident\_zip} zip codes should 
validate against the USPS database, which contains 
44,173 valid zip codes. We discovered that the \texttt{incident\_zip} 
field has 0.07\% invalid entries, resulting in 4,374 errors. This includes
zip codes such as ``12345'', ``10000'', and ``99999''. If you group 
the invalid zip code entries by percentage by agency, it 
mirrors the \emph{overall} breakdown of SRs by Agency, potentially 
indicating a systemic problem.

\paragraph{Created and Closed Dates:}
\label{sec:negativeduration}
With the \texttt{created\_date} and \texttt{closed\_date} fields, one 
might expect that these fields would be automatically populated by the  
application software. Thus when creating  or closing an SR 
the the software would automatically populate these fields. 
Unfortunately, this does not seem to be the case, as several anomalies 
exist in the \texttt{created\_date} and \texttt{closed\_date} fields that
an automated process would normally exclude, including:
\begin{itemize}[left=1.5em]
\item 12,450 SRs with a \texttt{closed\_date} before
  \texttt{created\_date} creating a nonsensical ``negative duration''. 
\item Eight SRs with \texttt{created\_date(s)} or \texttt{closed\_date(s)} in 
  the distant past, specifically  \texttt{1900-01-01}. Despite few of 
  these entries, such extreme values can greatly distort analytical efforts. 
\item 163,720 SRs with identical \texttt{created\_date(s)} and \texttt{closed\_date(s)} 
  to the second, creating a nonsensical ``zero duration''. 
\item 247,000 SRs closed and/or created exactly at midnight, to the second.
\end{itemize}

\paragraph{Closed before Created (Negative Duration):}
Citizens, NYC Government Officials, and Agencies use the \texttt{created\_date}(s) and 
\texttt{closed\_date}(s) to measure the \emph{duration} of SRs, 
a useful surrogate to measure an agency's responsiveness. 
While duration is not directly present in the dataset, 
it is easily computed as the difference between
\texttt{closed\_date} and \texttt{created\_date}.  Duration is one of 
most frequently used 311 performance metrics 
and is frequently analyzed, such as determining if 
one NYC borough receives faster Agency responses than 
another borough for certain complaints. There are 
12,450 SRs where the \texttt{closed\_date} precedes the 
\texttt{created\_date}, generating nonsensical ``negative durations.'' 
Although this represents only 0.2\% of the dataset, these errors can 
significantly impact response time analyses. Eight SRs with extremely 
large negative durations, all originating from the 
Department of Homeless Services (DHS); all contain an entry 
of \texttt{1900-01-01} as the \texttt{closed\_date}. This results 
in very large negative durations exceeding $-$44,601 
days (or 122 years). Many other SRs have negative durations 
in excess of $-$300+ days. Such anomalies, though rare, can 
significantly skew statistical results if not addressed during 
data cleansing. Accordingly, these SR rows are removed from our analysis.

\begin{figure}[tbp]
  \centering
  \begin{subfigure}[t]{0.495\textwidth} 
    \centering
    \includegraphics[width=\textwidth]{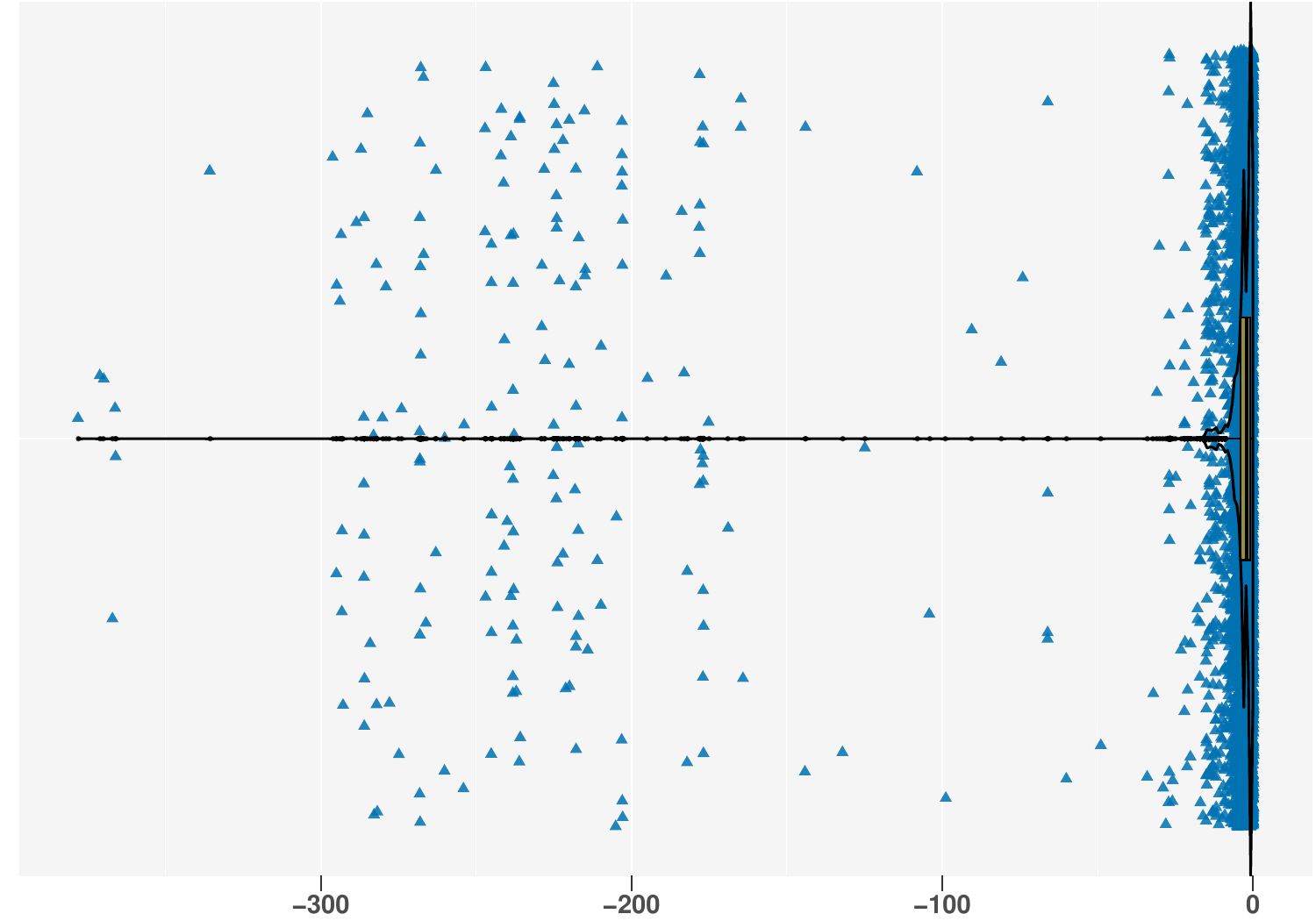}
    \caption{Distribution of SR durations.}
    \label{fig:negative-duration-violin}
  \end{subfigure}
  \hfill 
  \begin{subfigure}[t]{0.495\textwidth} 
    \centering
    \includegraphics[width=\textwidth]{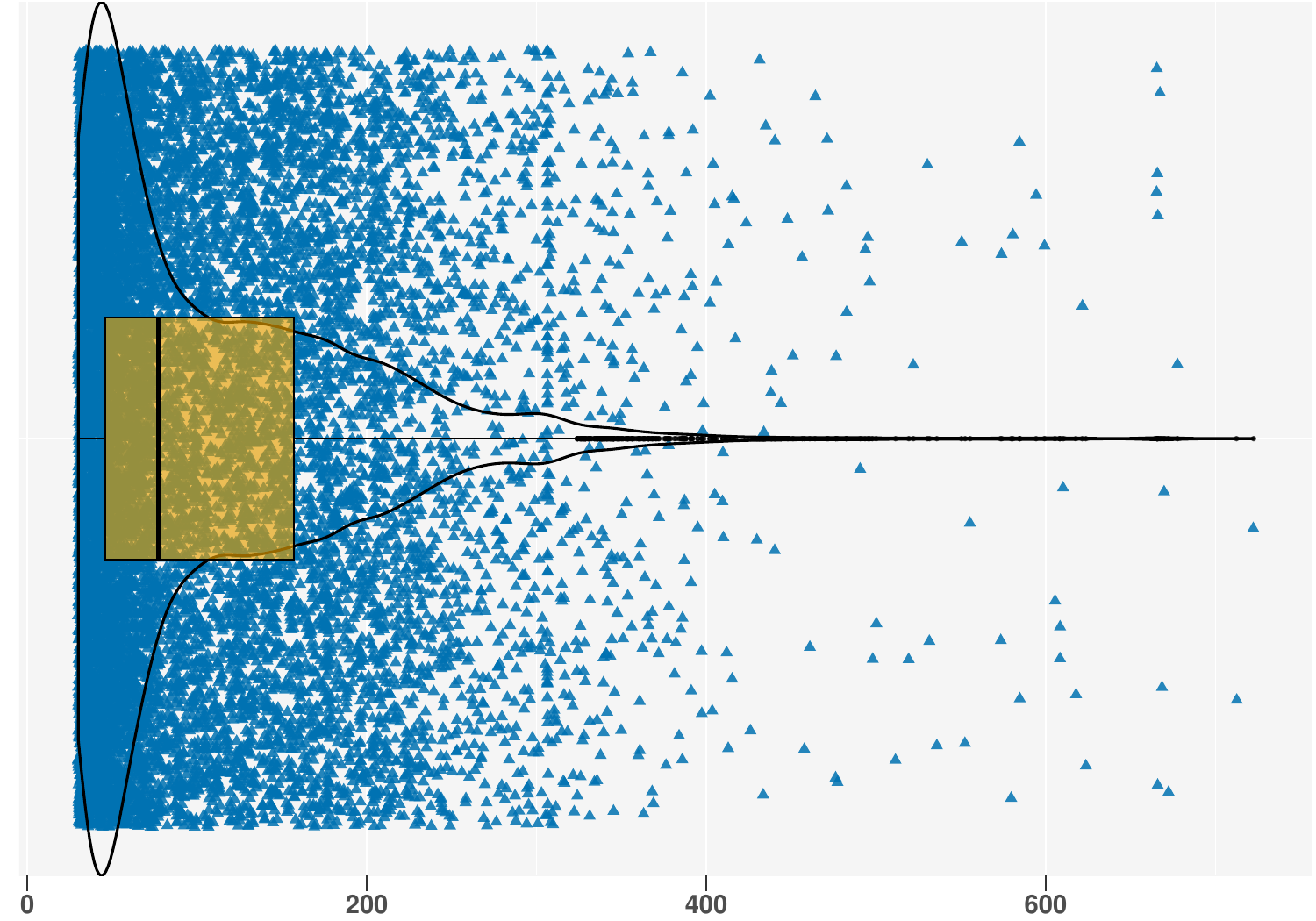}
    \caption{Post-closed \texttt{resolution\_action\_update\_date}(s) >30 days}
    \label{fig:resolution-violin}
  \end{subfigure}
  \caption{Violin plots for negative SR durations and post-closed
    resolution updates (2022--2023.}
  \label{fig:violin-plots}
\end{figure}

Excluding the extreme negative values (any that are lower than $-$730 days), 
Figure~\ref{fig:negative-duration-violin} shows the broad spread of 
negative duration SRs. While there are few outliers, the magnitude 
of these negative durations (some lower than $-$300 days) is troubling 
and can produce bizarre analytical results such as greatly 
distorting the mean response time for a specific complaint category.

Table~\ref{tab:largest-errors} summarizes the 
negative durations of the largest magnitude (in days). The data 
suggests that the negative duration issue is predominantly a problem 
within the Department of Transportation (DOT), where 95\% of these 
errors occur.

\begin{table}[tbp]
  \centering
  \caption{Sample  negative durations [excluding extreme values that
    are lower than $-$730 days]}
   \label{tab:largest-errors}
 	\begin{tabular}{l l l r l}
    \toprule
    \texttt{created\_date} & \texttt{closed\_date} & {Duration (days)}  & {Agency} \\
    \midrule
    2023-01-27 14:40:00 & 2022-01-14 14:40:00 & $-$378 & DOT \\
    2023-01-18 10:06:00 & 2022-01-12 10:06:00 & $-$371 & DOT \\
    2023-01-27 14:36:00 & 2022-01-22 14:35:00 & $-$370 & DOT \\
    2023-01-11 11:10:00 & 2022-01-09 11:10:00 & $-$367 & DOT \\
    2023-12-18 03:13:00 & 2023-01-16 13:10:00 & $-$335 & DOT \\
    \bottomrule
    \end{tabular}
 \end{table}

\paragraph{Identical Created \& Closed Dates (Zero Durations):}
A more prevalent issue occurs when the \texttt{closed\_date} and 
\texttt{created\_date} have exactly the same date-time stamp, 
to the second. This creates a \emph{zero duration} for the 
SR, which is nonsensical. There are 163,720 such zero duration 
SRs, representing 2.6\% of all non-blank data. 
Additionally, 99\% percent of these zero duration SRs 
occur within five agencies: Department of Health \& Mental 
Hygiene (DHMH), DOT, Department of Buildings (DOB), DSNY, and 
DEP. This distribution does not mirror the overall SR 
distribution by agency, suggesting an Agency-specific issue.
	
\begin{figure}[tbp]
    \centering
    \begin{subfigure}[t]{0.495\textwidth} 
        \centering
        \includegraphics[width=\textwidth]{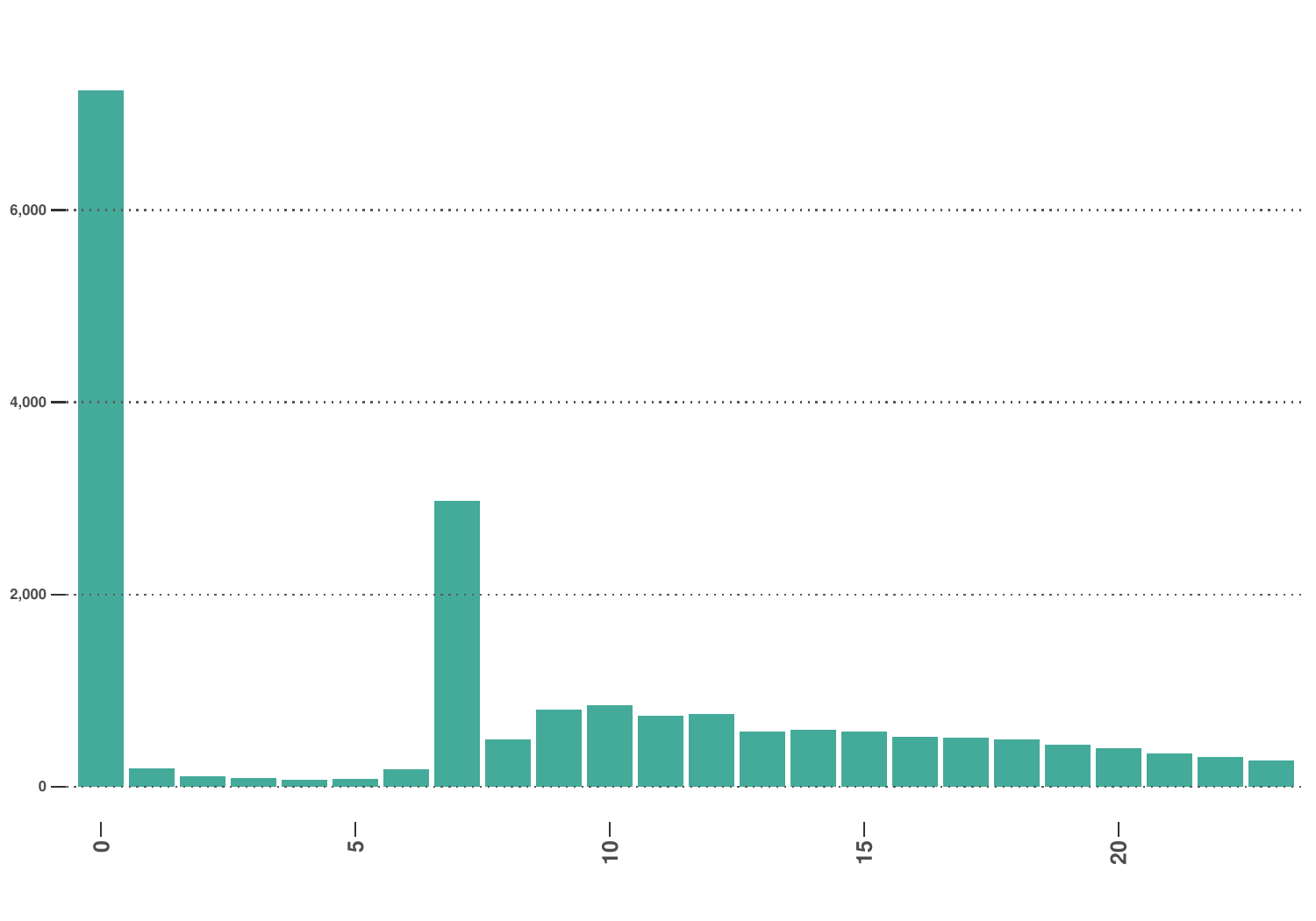}
        \caption{SRs created exactly on the hour}
        \label{fig:busiestcreated}
    \end{subfigure}
    \hfill 
    \begin{subfigure}[t]{0.495\textwidth} 
        \centering
        \includegraphics[width=\textwidth]{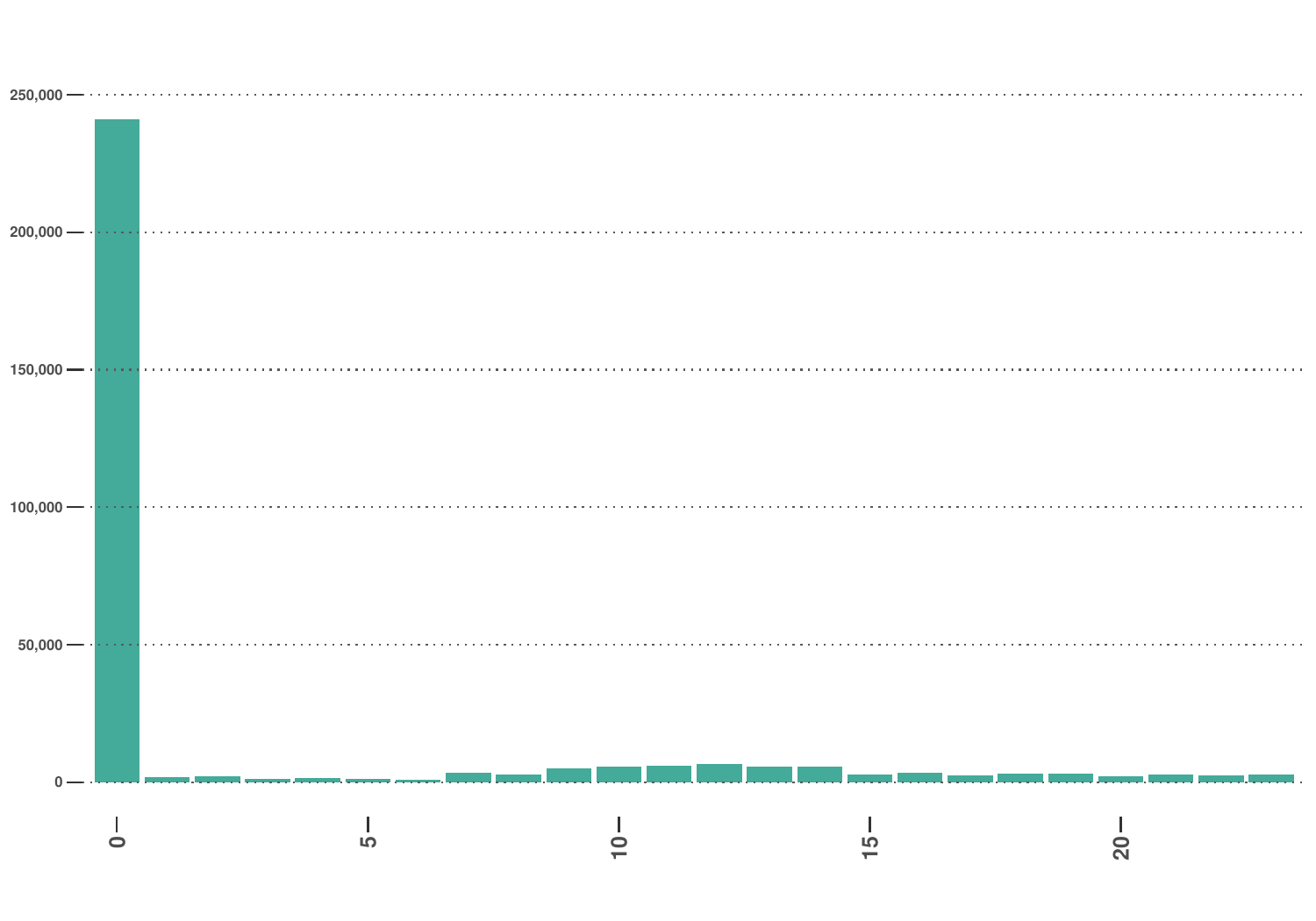}
        \caption{SRs closed exactly on the hour}
        \label{fig:busiestclosed}
    \end{subfigure}
    \caption{311 SRs Created \& Closed exactly on the hour (hh:00:00) for 2022--2023.}
    \label{fig:exacthours}
\end{figure}

\paragraph{Created or Closed at Midnight:}
Another issue involves the unusually large number of 
SRs where the \texttt{created\_date} and \texttt{closed\_date} 
indicate action exactly at midnight (00:00:00), 
to the second. Normally, SR creation and closure follows 
the work-day schedule, with most SRs created 
during working hours and fewer at night or in the 
early morning. However, there is a significantly higher 
number of SRs closed exactly at midnight, as well 
as a large number created exactly at midnight. The magnitude 
of this anomaly is well outside the 3$\sigma$ level as shown 
in Figure~\ref{fig:exacthours}

These unusual patterns of SR creation/closure at midnight 
suggest the presence of a bulk create/close software process, perhaps one
that automatically assigns time stamps of midnight (00:00:00) to 
 batches of SRs prior to sending the data to the 311 
database. This behavior distorts the calculated duration of these 
SRs as well as overall system usage by providing inaccurate 
dates. Further analysis of the distribution by Agency shows that over 92\% of 
the ``closed-exactly-at-midnight'' SRs come from just two 
Agencies: DOB (69\%) and DSNY  (23\%).

\paragraph{Resolution Action Update Date}
A similar issue occurs when an SR is modified (and the \texttt{resolution\_action\_update\_date} is altered) well 
after the \texttt{closed\_date}.  When an SR is 
updated, the 311 software automatically populates the \texttt{resolution\_action\_update\_date}. While it is certainly 
possible, and even routine, to update an SR after it is closed, 
some of these updates seem to be well beyond any 
normal or expected timeframe, with some ``updates'' 
occurring over 2 years later. As shown in Figure~\ref{fig:resolution-violin}, 
there are 10,265 SRs that were updated $>30$ days after the 
\texttt{closed\_date} (Not that Figure~\ref{fig:resolution-violin} excludes 
values $>730$ days, an arbitrary cutoff to exclude 
infeasible dates such as \texttt{1900-01-01}. The chart 
highlights the distribution of these post-closure 
 updates, with a notable concentration of updates occurring within 
the 30--200 day range. This pattern raises 
questions about whether such delayed updates are standard 
operational practice or indicative of a potential issue 
which warrants further investigation. It should be noted
that 84\% of these late updates are associated with the TLC agency
and 12\% with DSNY.

\subsection{Accuracy and Precision}
\label{sec:precision}
The question of precision vs. accuracy arises with the \texttt{latitude} 
and \texttt{longitude} fields. as well as with the duplicative 
\texttt{location} field. These fields are expressed as
a 14-decimal number, e.g. \texttt{latitude} 40.86769186022511. 
Given that 1 degree of latitude at the equator is equal to 111.04 
kilometers, the ``1'' at the end of that decimal number represents 
approximately 1.1 nanometers (1/1,000,000,000 of a meter). (For 
reference a DNA molecule is approximately 2 nanometers in width.) This 
representation appears to be a classic case of extreme precision, albeit with limited 
accuracy. All the \texttt{latitude}, \texttt{longitude}, and \texttt{location}
fields display this characteristic.

\subsection{Redundant Fields}
\label{sec:duplicates}
This analysis revealed several redundant fields, indicting a potential need 
for consolidation.

\paragraph{\texttt{latitude/longitude} \& \texttt{location}:} 
The \texttt{location} field is a pure concatenation of 
the \texttt{latitude} and \texttt{longitude} fields albeit with a 
comma and parentheses added. Arguably, this makes the \texttt{location} field 
more difficult to use than the individual fields. For example: 

\begin{center}
\begin{tikzpicture}[node distance=2cm and 1.5cm, >=stealth]

    \node (latitude) at (-1.5, 0.9) {40.768456429488};
    \node (longitude) at (1.8, 0.95) {$-$73.9575661888774};
    \node (location) at (0.285, 0) {(40.768456429488, $-$73.9575661888774)};

    \node[anchor=west] at (-8.5, 0.9) {\texttt{latitude} \& \texttt{longitude} fields};
    \node[anchor=west] at (-8.5, 0) {\texttt{location} field};

    \draw[->] (latitude.south) -- ([xshift=-1.77cm]location.north);
    \draw[->] (longitude.south) -- ([xshift=1.525cm]location.north);

\end{tikzpicture}
\end{center}

\paragraph{\texttt{borough} \& \texttt{park\_borough}:} These two fields are fully redundant, 
as they are 100\% matches.

\paragraph{\texttt{borough} \& \texttt{taxi\_company\_borough}:} Despite 
their names, these fields are almost entirely different, with only 
a 0.05\% match. The \texttt{taxi\_company\_borough} field is 
used exclusively by the TLC, indicating a need for consultation 
with that agency to understand the differing uses. It is 
likely that that the usage of \texttt{taxi\_company\_borough} is 
quite different than that of \texttt{borough}. 

\paragraph{\texttt{agency} \& \texttt{agency\_name}:} The \texttt{agency} 
field contains the abbreviations of City agencies (e.g., NYPD, DOT, DSNY), 
while \texttt{agency\_name} contains the full names. Including both seems 
redundant, especially given the space taken up by full names in the 
CSV file and the fact that the agency abbreviations are 
easily recognized and well understood.

\paragraph{\texttt{landmark} \& \texttt{street\_name}:} Per the 
311 SR Data Dictionary, the \texttt{landmark} 
field is intended to refer to any noteworthy location, 
including but not limited to parks, hospitals, airports, sports facilities, 
and performance spaces. However, most entries are actually 
street names, with a 62\% match to the \texttt{street\_name} field. 
Many of the non-matches (excluding blanks) also appear to be 
logical matches, differing only in spelling or formatting 
(e.g., ``NINTH AVE'' vs. ``9 AVE''). This indicates a possible 
misuse of the \texttt{landmark} field and highlights significant
redundancy. 

\paragraph{\texttt{cross\_street\_1 \& 2} \& \texttt{intersection\_street\_1 \& 2}:} These 
two street pairs are used to help identify the 
\texttt{incident address}. We found 88\% of both of these two 
street pairs (1 \& 2) to be duplicates. Such a high level of duplication 
raises the question of which field-pair should be trusted; 
and if these street pairs are used differently, how should they be treated?
It is possible that these two street pairs originated from different 
agencies that use different variable names to represent the same 
information, but no documentation exists to clarify this 
relationship. Nonetheless, these two street-pairs are highly redundant.

We attempted to standardize street names to reconcile entries that
were clearly identical but differed due to spelling or naming
conventions. For example, we sought to match entries such as
\texttt{KANE PL} and \texttt{KANE PLACE} or \texttt{AVE P} and
\texttt{AVENUE P}. However, this standardization effort was
ultimately abandoned, as it resulted in less than a 1\% increase in
matched street names.

\subsection{Reducing File Size}
\label{sec:filesize}
By removing duplicate and near-duplicate fields, one can reduce file size by 39.5\%, 
equating to a reduction of 1.4GB, a significant amount. A smaller file size 
means faster downloads, reduced storage impact, and often simpler data analysis. 
Below is a list of duplicate and near-duplicate fields which we believe could be
removed from the dataset with minimal data loss. 

\begin{itemize}[left=1.5em]
    \item \texttt{agency\_name}: Since each row contains the \texttt{agency} 
    field, which uses clear and well understood abbreviations, we believe the
    \texttt{agency\_name} can be removed with minimal data loss.
    
    \item \texttt{park\_borough}: This field is a 100\% match with 
    the \texttt{borough} field. It can be deleted without data loss.
    
    \item \texttt{location}: This field is a straight concatenation of 
    the \texttt{latitude} and \texttt{longitude} fields. It can be 
    deleted without data loss.
     
    \item \texttt{cross\_street\_1/2} and \texttt{intersection\_street\_1/2}: 
    These two street pairs exhibit an 88\% match. Although we cannot 
    confirm the accuracy of either of these pairs, we recommend deleting the 
    \texttt{intersection\_street\_1/2} fields, acknowledging 
    the concomitant data loss.
\end{itemize}

Additionally, certain fields may not be useful for broad analysis
due to their sparse data population. This applies to the fields 
below which are 99\% (or greater) blank. 
\begin{itemize}[left=1.5em]
    \item \texttt{taxi\_company\_borough}
    \item \texttt{road\_ramp}
    \item \texttt{vehicle\_type}
    \item \texttt{due\_date}
    \item \texttt{bridge\_highway\_direction}
    \item \texttt{bridge\_highway\_name}
    \item \texttt{bridge\_highway\_segment}
    \item \texttt{taxi\_pick\_up\_location}
\end{itemize}
While such sparsity may limit their use in 
broad analyses, these fields can still contain valuable insights 
for specific agency-level studies, given that even a 1\% 
non-blank rate equates to approximately 64,000 rows in this 
6.4 million-row dataset. Proper handling of sparse fields, 
such as storing them in separate tables or using them in 
targeted analyses, could improve data accessibility and 
efficiency without compromising the dataset’s broader utility.

\paragraph{Field Encoding:} Further size reduction is possible 
by encoding selected categorical variables 
Using \texttt{complaint\_type} as an example,
instead of repeatedly storing the full text of each possible value,
one could use a numeric representation. For example, 
``\texttt{NOISE - STREET/SIDEWALK}'' could be represented by a 
numeric code. For this dataset this code occurs 
over 300,000 times. Encoding not only reduces file 
size, but would likely improve processing efficiency by using numeric
values instead of text strings. Moreover, standardized 
categorical inputs often prevents misspellings 
and invalid entries, ensuring cleaner, more reliable 
data. This approach is particularly useful in large datasets, where 
frequently repeated text-based categorical variables significantly 
increases storage requirements.

One promising alternative to reducing data storage  
for this large 311 SR dataset (3.7GB) would be employment of 
the cross-platform Apache Arrow package. Apache Arrow (arrow.apache.org) 
is an innovative, cross-language development platform designed for 
in-memory data. It defines a standard for representing columnar data, 
enabling efficient analytics and real-time data processing. As such,
it offers dramatic improvements in file storage size, as well as 
efficient analysis and cross-platform support for R, Python, 
and Julia programming languages. Remarkable storage size 
reduction (within several orders of magnitude) is possible \citep{bates2024csv}.

\subsection{Data Dictionary} 
\label{sec:datadictionary}

The 311 SR Data Dictionary needs updating. We found several 
discrepancies between the Data Dictionary and the actual
data. Such errors could potentially mislead analytic efforts. Examples::

\paragraph{Data typing:} While the NYC Open Data portal displays the data types of 
various fields (aka ``columns''), the Data Dictionary does 
not. For example, in the Data Dictionary, 
the \texttt{incident\_zip} field is specified as ``text.'' This may 
not be the most suitable data definition. Perhaps a better 
dictionary description would be  ``categorical'' with a note 
indicating that the data must contain only numeric values 
and is not subjected to arithmetic operations.

\paragraph{Domain of Legal Values:} Many of the data fields have
domains of legal values that are missing, incomplete, or inaccurate. For 
example, the Data Dictionary indicates 
that the \texttt{status} field takes on values of \texttt{assigned, canceled, 
closed}, or \texttt{pending}. However, we observed additional values 
such as \texttt{in progress, started}, and \texttt{unspecified}. Additionally, 
no SRs were found with a status of \texttt{canceled}. Similarly, 
the Data Dictionary for the \texttt{address\_type} field lists as acceptable
values \texttt{address, blockface, intersection, latlong}, and 
\texttt{placename}. However, we found additional values such 
as \texttt{bbl} and \texttt{unrecognized}. Additionally, no 
\texttt{latlong} values were observed. Other fields, such 
as \texttt{facility\_type}, \texttt{vehicle\_type}, \texttt{taxi\_pick\_up\_location},
texttt{road\_ramp}, and \texttt{city}, exhibit similar inaccuracies 
in their specified domain values.

\paragraph{Missing data fields:} As mentioned earlier, six \texttt{@computed} 
fields that are present in the data but not included in 
the Data Dictionary need to be documented and perhaps 
labeled as ``experimental'' or ``not for official use''.

\section{Government Open Data Curation Recommendations}
\label{sec:recommendations}
Building upon  insights from our case study of the 
NYC 311 SR data, we propose a set of data 
curation principles tailored for government open datasets. 
These principles are designed to address the unique challenges 
 observed in the curation of such datasets, 
ensuring they remain reliable, consistent, and useful for 
public services, research, and other applications.
To facilitate implementation, the principles are prioritized based on
their impact and feasibility, offering a clear roadmap for improving
data curation practices.

\subsection{Principles}

\paragraph{Principle 1: Establishing Cross-Organizational Consistency}
A key challenge in government released datasets is the 
harmonization of data fields across various Agencies, and 
maintaining that consistency over time. Inconsistent field
definitions, data formats, or value domains can lead to erroneous
analyses and hinder cross-agency collaboration.
To address this, Agencies should adopt 
standardized naming conventions, value domains, usages, and 
formats across all datasets. Regular audits and data harmonization 
processes should be implemented to ensure consistency 
across long-term datasets, particularly as Agency structures 
and reporting standards evolve. For instance, ensuring that 
historical datasets reflect the same field naming conventions 
over a span of years can prevent inconsistent analyses. 
These standards should be promulgated by the Open Data 
governing authority, following agreements by the relevant parties.

\paragraph{Principle 2: Ensuring Data Entry Accuracy}
Data accuracy and validity are fundamental to the utility of 
open datasets. In the NYC 311 data, issues such as 
invalid zip codes and nonsensical date entries 
highlight the need for rigorous validation processes. Governments 
should establish clear protocols for data entry, ensuring that 
fields are populated with valid, legally acceptable values. Accurate
data entry reduces the risk of introducing systemic errors,
particularly in high-volume datasets. To
the maximum extent possible,the software should assist in 
ensuring data entries are accurate and valid. Some examples include:

\begin{itemize}[left=1.5em]
    \item Certain date fields should be promulgated by the application software
     versus manual entry. Dates should be populated
    based upon an action in the system, such as a change in status. 
       
    \item Select fields should be subjected to automated checks 
    that prevent logical errors, such as a closing date occurring before the creation date. 
    
    \item A number of fields have clearly defined domains of legal
    values. These fields should be validated against a reference 
    dataset upon entry. 
 \end{itemize}

\paragraph{Principle 3: Optimizing Storage Efficiency}
Efficient storage and data representation are crucial for handling 
large government datasets. Our research found several duplicate and
near-duplicate fields in the 311 SR dataset. Removing these
redundancies could reduce the dataset size by 39.5\%, improving
performance without compromising analytical value. Eliminating 
redundant fields and encoding categorical variables can 
significantly reduce file sizes and improve the efficiency 
of data analysis. Similarly, categorical variables could be 
encoded in a standardized format to optimize storage and 
streamline analyses. Reducing file size not only saves 
storage space but also enhances the performance of queries, 
downloads,  and analysis processes.

\paragraph{Principle 4: Updating Data Dictionaries}
Clear and accurate data documentation is essential for the effective use 
of government datasets. Our study found several 
instances where the Data Dictionary did not accurately reflect the 
actual data. For example, undocumented fields and outdated domain
definitions in the 311 SR dataset introduced ambiguity, complicating
analysis. Domains of legal/allowable values should be referenced
in the Data Dictionary. Regular reviews and updates should 
occur whenever there are changes in data structure, data usage, or 
content. This practice enhances the ability of external users, 
including researchers and developers, to work 
effectively with the data.

\paragraph{Principle 5: Automating Quality Assurance Processes}
Automating data validation and quality assurance processes can 
significantly improve the reliability of government datasets. Certain
unusual trends can be identified, and irregularities highlighted. 
In the NYC 311 case, issues such as large spikes in SR 
creation and closure at exactly midnight (or noon) likely 
stem from a cron script process that distorts the 
accurate capture of SR durations. Real-time validation systems could
prevent such errors, ensuring data integrity at the point of entry. By
implementing
real-time and event-driven validation tools, governments 
can both prevent such errors from entering the dataset in the first 
place, as well as highlight anomalous trends. Automated 
systems can be designed to flag inconsistencies, incorrect values, 
illogical sequences, or illogical data entries. Automation can 
also facilitate faster detection and correction of errors, reducing 
the need for manual intervention and ensuring the data 
remains clean and reliable as it is continuously updated.

\paragraph{Principle 6: Addressing Data Transparency}
Transparency is a core tenet of open data, and governments must ensure 
that datasets are not only accurate but also easily accessible and 
readily understandable. This involves providing clear 
metadata and making datasets available in user-friendly 
formats. This includes publishing clear metadata, which enables users
to understand the dataset’s structure, limitations, and intended use
cases. Additionally, governments should communicate transparently
about data quality issues, such as notifying the public when errors in SR 
fields (e.g., erroneous \texttt{closed\_ date(s)} are discovered, including
an explanation about how such issues will be resolved. By 
ensuring transparency and keeping the public informed, 
governments foster trust and promote wider use of datasets 
for research, civic engagement, and innovation.

\subsection{Actionable Steps}
While these principles are broad, actionable steps are required to 
implement them effectively. Prioritizing these steps will help
governments address the most critical data curation challenges
first. Governments can begin by investing in
real-time validation systems, regular updates to data dictionaries, 
and tools for encoding and standardizing data fields. This would include
systems that automate enforcement of data validation rules 
(e.g., flagging incorrect zip codes), significantly improving data 
quality. Automated systems can proactively detect and correct common
issues, reducing the manual workload.
Collaboration with city agencies and stakeholders is crucial
to ensure that data standards, formats, and metadata meet the needs 
of diverse users, including policymakers, researchers, and the 
general public. Regular communication between agencies and external 
users can help identify recurring issues and areas where improvements 
are needed. By facilitating regular communication and collaboration 
with various stakeholders, governments can ensure data remains relevant 
and actionable for public use.

Automated Quality Assurance techniques should be employed to 
monitor the quality of the data, as measured by agreed upon
standards, such as the percentage of invalid zip codes. Such tools could
capture and display the quality of the data sets as presented
on appropriate dashboards. These dashboards provide a real-time
snapshot of data quality, making deviations immediately visible and
actionable. Deviations in quality would be
automatically highlighted, alerting appropriate corrective actions.

Engagement with data scientists and statisticians is key to improving 
data curation strategies. These experts can help design and refine 
curation processes, such as creating algorithms for automatic data 
cleaning, identifying inconsistent data entries, and ensuring the 
dataset(s) structure is maintained over time. Their involvement also
helps to establish best practices for validating incoming data and
maintaining consistency across data sources.
Feedback, such as the issues reported 
in this paper, benefit from a clear and accessible channel to the 311 SR data 
team and other relevant City Agencies. Establishing such feedback loops 
and other opportunities for public engagement, ensures that 
recommendations for improving data curation and validation 
are not only heard and discussed, but implemented. An iterative
feedback system enables timely resolution of issues and fosters
long-term improvements in data management.

An especially valuable engagement avenue is through events such as NYC 
Open Data Week, which foster collaboration and innovation across sectors. 
Our involvement in this project stemmed from such an event, where we were 
inspired by the potential of open data to enhance public services. Governments 
should continue these types of initiatives, which not only promote 
data literacy, but encourage community-driven improvements to
datasets. These events can strengthen public engagement and
demonstrate the societal value of open data initiative
By leveraging the expertise of data scientists, statisticians, stakeholders, 
and civic organizations, governments can ensure their open datasets 
remain accurate, transparent, and useful for a wide range of applications.

\section{Discussion} 
\label{sec:discussion}
Our study highlights the critical role of robust data curation in
ensuring the reliability and utility of open datasets, as demonstrated
through the examination of the NYC 311 data. As these datasets are
applied across diverse fields such as public services and academic
research, inconsistencies, missing values, and formatting errors can
significantly undermine the insights derived from them. Trustworthy
machine learning systems, which rely heavily on high quality data, are
especially vulnerable to such issues. Errors in the underlying data,
such as biases or inconsistencies, can compromise both the accuracy
and fairness of predictive models, impacting real-world decisions in
urban planning and policy \citep{rahm2000data, geiger2020garbage}.

While the findings here are specific to NYC 311 data, the principles 
proposed have broader applicability. Datasets in healthcare, 
transportation, and environmental monitoring can similarly benefit 
from these curation strategies, enabling improved outcomes across 
domains.
This research further emphasizes the necessity of harmonizing data
across City agencies, particularly resolving discrepancies in field
names, formats, usage, and definitions. Consistency in data is essential for
trustworthy machine learning, where inconsistencies across sources can
lead to distorted outcomes. This challenge is even more pronounced in
datasets that span long time periods, as changes in agency structures
or reporting standards which can introduce subtle biases and data 
distortions. Long-term data consistency is crucial for 
longitudinal studies and predictive modeling, as even minor 
data shifts can lead to significant deviations in model outcomes, 
as noted by \citet{rahm2000data} and\citet{borgman2012conundrum}.

Finally, the intersection of data curation and machine learning for
public policy applications opens new avenues for improving governance.
High quality data enables machine learning models to better predict
trends, allocate resources, and address service delivery issues, such
as delays in responding to 311 complaints. The COVID-19 pandemic
underscored the importance of real-time, trustworthy data in crisis
management, where the accuracy and timeliness of data-driven insights
were critical for public health responses. Without proper data
curation, machine learning models used during the pandemic would have
been compromised, affecting decisions on resource allocation and
service provision \citep{worby2020face, khemasuwan2021applications}.
Future efforts could focus on automating data curation processes,
particularly in real-time data pipelines, to ensure that the data used
in machine learning models remains accurate, clean, and reliable
\citep{chu2016data, hurbean2021open}.

\section*{Supplementary Material}
The NYC 311 R code used to produce these results, along 
with additional analytical findings, are available 
at \url{https://github.com/jun-yan/nyc311clean}. Reference 
data files (311 SR data of 2022--2023, USPS zip codes) are available 
at \url{https://doi.org/10.6084/m9.figshare.c.7617311}.

\bibliographystyle{jds}
\bibliography{refs}

\end{document}